\documentclass[journal]{IEEEtran}
\usepackage{amsmath,amsfonts}
\usepackage{algorithmic}
\usepackage{algorithm}
\usepackage{array}
\usepackage[caption=false,font=footnotesize,labelfont=rm,textfont=rm]{subfig}
\usepackage{textcomp}
\usepackage{stfloats}
\usepackage{url}
\usepackage{verbatim}
\usepackage{graphicx}
\usepackage{cite}
\usepackage{subfig}
\usepackage{cleveref}

\makeatletter
\usepackage{xcolor}
\newcommand{\rev}{\textcolor{blue}}

\renewcommand{\maketag@@@}[1]{\hbox{\m@th\normalsize\normalfont#1}}%

\makeatother

\begin{document}

\title{Holographic Communication via Recordable and Reconfigurable Metasurface }

\author{Jinzhe Wang,
Qinghua~Guo,~\IEEEmembership{Senior Member,~IEEE,} 
and~Xiaojun~Yuan,~\IEEEmembership{Senior Member,~IEEE}

\thanks{Jinzhe Wang and Xiaojun Yuan are with the National Key Laboratory of Wireless Communications, University of Electronic Science and Technology of China, Chengdu 611731, China (e-mail: wangjinzhe@std.uestc.edu.cn; xjyuan @uestc.edu.cn).

Qinghua Guo is with the School of Electrical, Computer and Telecommunications Engineering, University of Wollongong, Wollongong, NSW 2522, Australia (e-mail: qguo@uow.edu.au).}}



\maketitle
\begin{abstract}    
Holographic surface based communication technologies are anticipated to play a significant role in the next generation of wireless networks. The existing reconfigurable holographic surface (RHS)-based scheme only utilizes the reconstruction process of the holographic principle for beamforming, where the channel state information (CSI) is needed. However, channel estimation for CSI acquirement is a challenging task in metasurface based communications. 
In this study, 
inspired by both the recording and reconstruction processes of holography, we develop a novel holographic communication scheme 
by introducing recordable and reconfigurable metasurfaces (RRMs), where channel estimation is not needed thanks to the recording process. Then we analyze input-output mutual information and \rev{outage probability} of the RRM-based communication system and compare it with the existing RHS based system.
Our results show that, without channel estimation, the proposed scheme achieves performance comparable to that of the RHS scheme with perfect CSI, suggesting a promising alternative for future wireless communication networks.






\end{abstract}


\section{Introduction}
\IEEEPARstart{T}{he} upcoming {sixth generation (6G) wireless communication networks are set to revolutionize mobile connectivity and deliver high-throughput data services through energy-efficient, highly integrated and affordable communication infrastructure and devices \cite{ref1,ref2,ref3}. 
Recently, large antenna arrays with hundreds or even thousands of elements have received much attention, which enable the massive multiple-input multiple-output (MIMO) technology to boost the network capacity and communication reliability by harnessing multiplexing, diversity and power gains \cite{ref4}. However, conventional arrays face intrinsic limitations that prevent them from fully realizing the ambitious goals of 6G networks. 
Specifically, the beamforming gain of a conventional antenna array is contingent on various hardware components, including phase shifters and power amplifiers, essential for constructing complex phase-shifting circuits. This reliance results in high hardware costs and energy consumption, posing challenges when scaling up to larger configurations \cite{ref5}. Consequently, the development of innovative antenna manufacturing technologies has emerged as a pivotal research area in the realm of 6G advancement. Pursuing such technologies is essential to surpass current limitations and enable the full realization of 6G network capabilities.
}





{More recently, metamaterial-based surfaces have emerged as a promising solution to achieve large-scale array configurations in a cost- and energy-efficient manner. Metamaterials offer the capability to manipulate electromagnetic wave radiation by achieving arbitrary permittivity and permeability. Through strategic arrangements of meta-atoms, these materials can control spatial electromagnetic waves to attain desired amplitudes or phases. Metasurfaces introduce a paradigm of software-reconfigurable antennas owing to their programmability and tunability. Reconfigurable metasurfaces have emerged as a key enabler in 6G, which can operate in passive or active modes.} 
{Reconfigurable intelligent surfaces (RISs), comprising massive passive reflective elements, show great potentials in enhancing the performance of wireless networks by intelligently configuring propagation environments to achieve favorable channel responses \cite{ref21,ref26}. 
However, RISs, typically deployed between communication terminals, cannot function as transceivers. In contrast, reconfigurable holographic surfaces (RHSs), consisting of numerous metamaterial radiation elements, can actively generate beams in desired directions \cite{ref20}.} 

{The concept of holographic antennas was originally extended from leaky-wave antennas\cite{ref10}. Early holographic antennas utilized relatively simple hardware structures, e.g., the holographic artificial impedance surfaces \cite{ref11}, which manipulate electromagnetic waves by designing surface impedance. However, since impedance surfaces are not reconfigurable, their application has been limited. Subsequent research has focused on RHS, with reconfigurability achieved through various mechanisms, e.g., surfaces based on liquid crystals (LCs)\cite{ref12} and positive-intrinsic-negative (PIN) diodes\cite{ref13}. 
Due to their lightweight and low hardware costs, RHSs have been recognized as a key enabler in 6G. Companies like Pivotal Commwave \cite{ref15} and Kymeta \cite{ref16} have been involved in developing commercial RHS prototypes.} \rev{In parallel, research on photonic tightly coupled antenna (TCA)-based EM holography has enabled bidirectional RF–optical conversion using electro-optic modulators and photodetectors. This approach offers ultra-wide bandwidth, high efficiency, and real-time processing via spatial light modulators, though its high complexity and cost have limited practical use in wireless communications \cite{ref39}. In addition, due to the powerful EM wave control capabilities of large-scale holographic antennas, modeling in the EM domain is also an important area of research. For example, in \cite{ref32}, a generalized EM-domain near-field channel model is proposed, and the capacity limits in a Line-of-Sight (LoS) environment are studied.}
\IEEEpubidadjcol



{Holographic beamforming using RHSs, based on the reconstruction process of holography, has garnered significant attention. A holographic beamforming solution was proposed in \cite{ref17}, and its impact on system performance was investigated in \cite{ref18}, where an optimization scheme for multi-user beamforming was proposed.} The holographic-pattern division multiple access (HDMA) was developed in \cite{ref22} by mapping the expected signal of the receiver onto superimposed holographic patterns.  Subsequently, a multi-user holographic beamforming scheme for HDMA was proposed. 
Additionally, HDMA was applied to Low Earth Orbit (LEO) satellite communication scenarios in \cite{ref23}. Furthermore, based on the hardware design and full-wave analysis of RHS, an RHS prototype was implemented. With the prototype, an RHS-assisted communication platform was constructed to further validate the feasibility of holographic radio supported by RHS\cite{ref24}.  However, optimizing the beamformer in RHS-based holographic beamforming requires the knowledge of CSI obtained through channel estimation, which poses a significant challenge. RHS typically comprises a large number of antenna elements or even features a continuous aperture, making practical channel estimation a complex task, as a vast number of channel coefficients need to be estimated \cite{ref6,ref7,ref8}. 

{In this work, we address these challenges in RHS-based communications by avoiding channel estimation. We introduce recordable and reconfigurable metasurfaces (RRMs) and propose a RRM-based communication scheme inspired by both the recording and reconstruction processes of holography, which is termed as holographic communication\footnote{{Holographic communication in this work refers to communication based on the hologram principle. This differs from the holographic communication described in the literature, where data for 3D holographic imaging is transmitted through a communication system
\cite{ref19}.}}.} {RRMs differ from RHSs by incorporating a power recording module in the surface, enabling power recording of the interference between the local reference wave and the received signals from users.
The recorded interference power is then used for holographic beamforming for downlink transmission. We then analyze the mutual information of the RRM-based communication system,  demonstrating performance comparable to the RHS system with perfect CSI.}
The novelties and contributions of this work are summarized as follows:
\begin{itemize}

\item
{We develop a working principle for the novel RRM-based holographic communication scheme by incorporating both the recording and reconstruction processes of holography, eliminating the need for channel estimation thanks to the recording process.}

\item
\rev{We establish the relationship between the recorded interference power and the RRM holographic weights for downlink beamforming in time-division multiplexing systems. We further analyze the corresponding holographic beam pattern and apply lightweight computational optimizations to the weights to suppress sidelobe levels.}  

\item
\rev{We derive a baseband signal model for the RRM-based holographic communication system under a multipath delay-resolvable channel model, and analyze the mutual information between the input and output of the system. Additionally, we evaluate the outage probability based on the mutual information.}

\item
We compare the RRM-based system with RHS-based system, showing that the RRM-based system achieves performance close to that of the RHS-based system with perfect CSI. It is worth mentioning that the recording process in the RRM-based system can be implemented during uplink transmission, and channel estimation with pilot signals are no longer required. This greatly simplifies the communication system and reduces the communication overhead, providing a promising alternative solution.

Compared to the RHS-based system, the proposed RRM-based system does not require channel estimation and complex weight optimization for holographic beamforming.  

\end{itemize}

{The remainder of this paper is organized as follows. Section II introduces the principles of optical holography and electromagnetic (EM) holography proposed in this work. Section III details the proposed EM holography based on RRM and its working principle. Section IV elaborates the holographic communication system based on RRM, with a detailed derivation of the signal model and mutual information analysis. Section V provides simulation results, demonstrating the superiority of the proposed RRM-based system. Finally, Section VI concludes the paper.}

\textit{Notation}: 
Scalars are represented by italicized letters, vectors by bold italicized lowercase letters, and matrices by bold italicized uppercase letters. $*$ denotes convolution. For a complex scalar \( x \), its conjugate is denoted by \( \bar{x} \), its magnitude by \( |x| \), and its real and imaginary parts by \( \mathrm{Re}(x) \) and \( \mathrm{Im}(x) \), respectively. The \( i \)-th element of a vector \( \boldsymbol{v} \) is written as \( v(i) \). For a matrix \( \boldsymbol{M} \), \( M(i,j) \) denotes the element in the \( i \)-th row and \( j \)-th column. The notation \( \text{mean}(\boldsymbol{M}) \) refers to the average of the matrix elements. Additionally, \( \boldsymbol{M}^\mathrm{T} \) and \( \boldsymbol{M}^\mathrm{H} \) represent the transpose and conjugate transpose of \( \boldsymbol{M} \), respectively. Finally, \( \mathcal{CN}(0, \sigma^2) \) denotes a complex univariate Gaussian distribution with zero mean and variance \( \sigma^2 \).

\begin{figure*}[ht]
    \centering
    \subfloat[]{%
        \includegraphics[width=0.43\linewidth]{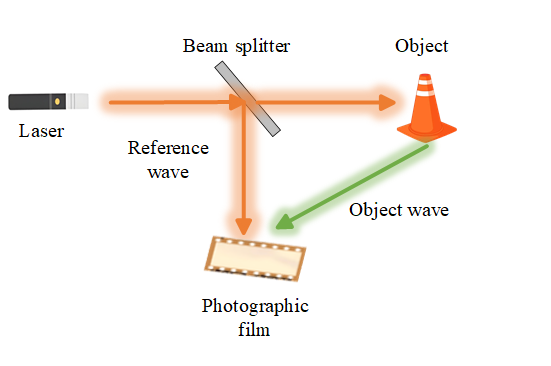}
        \label{fig1a}
    }
    \subfloat[]{%
        \includegraphics[width=0.43\linewidth]{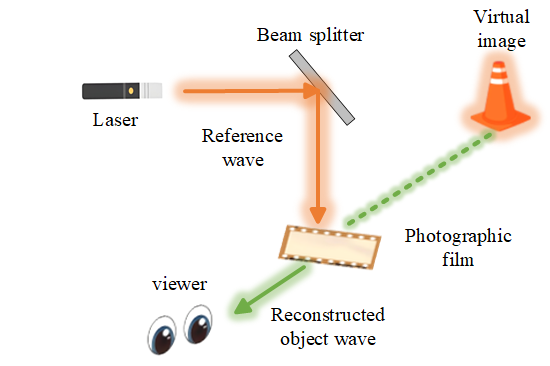}
        \label{fig1b}
    }\\
    \subfloat[]{%
        \includegraphics[width=0.43\linewidth]{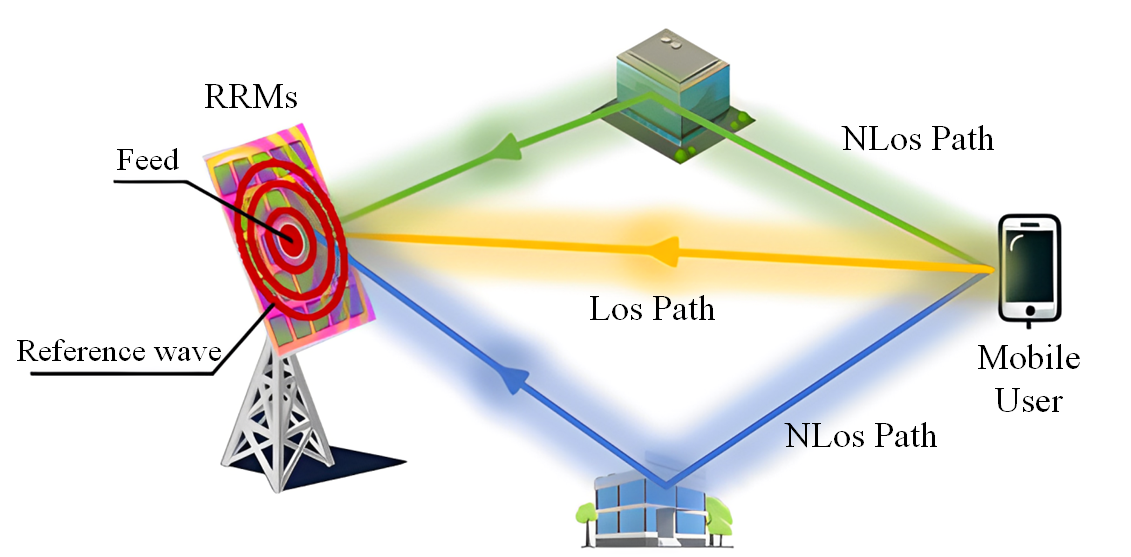}
        \label{fig1c}
    }
    \subfloat[]{%
        \includegraphics[width=0.43\linewidth]{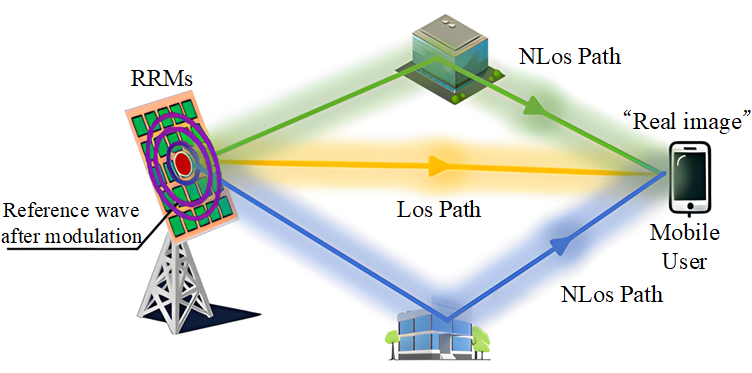}
        \label{fig1d}
    }
    \caption{Illustration of optical holography and EM holography: (a) Recording process of optical holography; (b) Reconstruction process of optical holography; (c) Recording process of EM holography; (d) Reconstruction process of EM holography.}
    \label{fig:2x2grid}
\end{figure*}

 \section{From Optical Holography to EM Holographic Beamforming}
 \label{sec:2}

In this section, we introduce the fundamental principles of optical holography, including both a recording process and a reconstruction process. These principles inspire the development of RRMs and our proposed electromagnetic (EM) holography.

{Optical holography is a technique for creating three-dimensional images using the principles of light interference and diffraction. By employing coherent light sources, such as lasers, holography captures the light wavefronts reflected from an object, encoding both the intensity and phase information into an interference pattern known as a hologram. This hologram can be recorded using photographic film or digital sensors. This recording process is shown in  Fig.1\subref{fig1a}. The reconstruction process, depicted in Fig.1\subref{fig1b}, involves illuminating the recorded hologram with a reference wave, which reconstructs the original light wavefronts. This process produces a three-dimensional image that can be viewed from different angles, offering realistic depth perception.} 


As shown in Fig. 1 (a), a laser is employed in the optical holography system to generate a reference wave. After passing through a beam splitter, the reference wave is divided into two beams. One beam of the reference wave directly illuminates the recording medium or device at position $\boldsymbol{r}$. The wavefront of this reference beam can be expressed as 
\begin{equation}
R\left( \boldsymbol{r}\right) =A_{R} e^{\mathrm{j} \phi _{R}(\boldsymbol{r}) },
\end{equation}
where $A_{R}$ and $\phi _{R}(r)$ represent the amplitude and phase of the reference wave, respectively. The other beam illuminates the target object, and the wave reflected from the target object is referred to as the object wave. The wavefront at position $r$ of the target object is represented as
\begin{equation}
O\left( \boldsymbol{r}\right) =A_{O} e^{\mathrm{j} {\phi _{O} (\boldsymbol{r})} },
\end{equation}
where $A_{O} $ and $\phi _{O} (\boldsymbol{r})$ respectively represent the amplitude and phase of the object wave. {The reference wave and the object wave meet at the recording medium or device, which is intensity sensitive. Then the interference intensity of the reference wave and the object wave is recorded, leading to the hologram that can be represented at position $\boldsymbol{r}$ as}
\begin{equation}
\begin{aligned}
{W\left( \boldsymbol{r}\right)} =&\left| O\left( \boldsymbol{r}\right) +R\left( \boldsymbol{r}\right) \right| ^{2}\\
=&\left| O\left( \boldsymbol{r}\right)\right| ^{2}+\left| R\left( \boldsymbol{r}\right)\right| ^{2}\\
&+R^{\ast }\left( \boldsymbol{r}\right)O\left( \boldsymbol{r}\right)+O^{\ast }\left( \boldsymbol{r}\right)R\left( \boldsymbol{r}\right),
\end{aligned}
\end{equation}
where the last two terms reflect the phase difference between the reference wave and the object wave. Therefore, the hologram not only records the intensity information of the two waves but also records their amplitude and phase information. 

{The holographic reconstruction process utilizes the hologram obtained during the recording phase. In this step, the laser emitter used in the recording process is employed to generate the reference wave, which is directed onto the hologram via a beam splitter. The intensity of the transmitted light field, excited by the reference wave, interacts with the hologram on the recording medium or device, producing the reconstruction wave}
\begin{equation}
    \begin{aligned}
    A_{rec}(\boldsymbol{r})=&W\left( \boldsymbol{r}\right) R\left( \boldsymbol{r}\right) \\
=&\left| O\left( \boldsymbol{r}\right)\right| ^{2}R\left( \boldsymbol{r}\right) +\left| R\left( \boldsymbol{r}\right)\right| ^{2}R\left( \boldsymbol{r}\right)\\ 
&+O\left( \boldsymbol{r}\right)\left| R\left( \boldsymbol{r}\right)\right| ^{2}+O^{\ast }\left( \boldsymbol{r}\right)R\left( \boldsymbol{r}\right)^{2},
\end{aligned}
\end{equation}
where the first two terms represent the waves formed along the direction of the reference wave, while the last two terms respectively denote the reconstructed object wave and the distorted conjugate object wave.

{Inspired by optical holography reconstruction processes, research on EM holographic communications has been carried out, exemplified by the development of the RHS technique. However, it requires the knowledge of CSI, which is challenging to acquire by channel estimation since a large number of channel coefficients need to be estimated. To circumvent this challenge, we develop a working principle of holographic communications by introducing RRM, which include both the recording process as shown in Fig. 1 (c) and holographic beamforming (reconstruction) process as shown in Fig. 1 (d). Thanks to the recording process, channel estimation is no longer needed. It is worth mentioning that the recording process can be implemented during the uplink transmission of data signals, avoiding the pilot overhead in a conventional communication system. 

\rev{However, directly applying optical holography to wireless communications is infeasible.  As shown in Fig. 1(b), the reconstruction process in optical holography forms a 'virtual image,' where the light rays do not actually converge but appear to originate from a point when extended backward. In contrast, wireless communication requires forming directed beams toward users, analogous to a 'real image' in optics, where the light rays actually converge at a specific point.  This fundamental distinction necessitates additional design of holographic beamforming weights following hologram recording.}

It turns out that our RRM-based holographic communication system can achieve performance close to that of the RHS-based system with perfect CSI, providing a promising alternative solution for future wireless networks.} 


\section{EM Holography Based on RRMs}
\label{sec:3}
In this section, we elaborate our RRM-based holographic beamforming and compare it with the RHS-based scheme in terms of hardware structure and implementations.
\rev{As shown in Fig. 2, RRM can be constructed using the leaky-wave antenna technique and comprises feeds, substrate and metamaterial radiation elements. The structure of RRM is similar to RHS but with one essential distinction: each metamaterial radiation element is integrated with a power sensor to measure and record the interference intensity between the reference wave and incoming user waves during the recording process. We adopt half-wavelength element spacing to prevent the power recorded at each unit from being affected by neighbouring elements. Then we propose a miniaturized probe-based power recording module in which one miniature probe is colocated with each antenna unit to sample the standing-wave power arising from the superposition of the reference and incident waves. The design of this module is supported by prior work on probe-insertion reflectometry and multi-port reflectometers, which demonstrate accurate waveguide-based power measurements while addressing probe-induced field perturbations through careful probe design and insertion-depth control \cite{ref28,ref29,ref30}.} 

\begin{figure}
        \centering
        \includegraphics[width=0.4\textwidth]{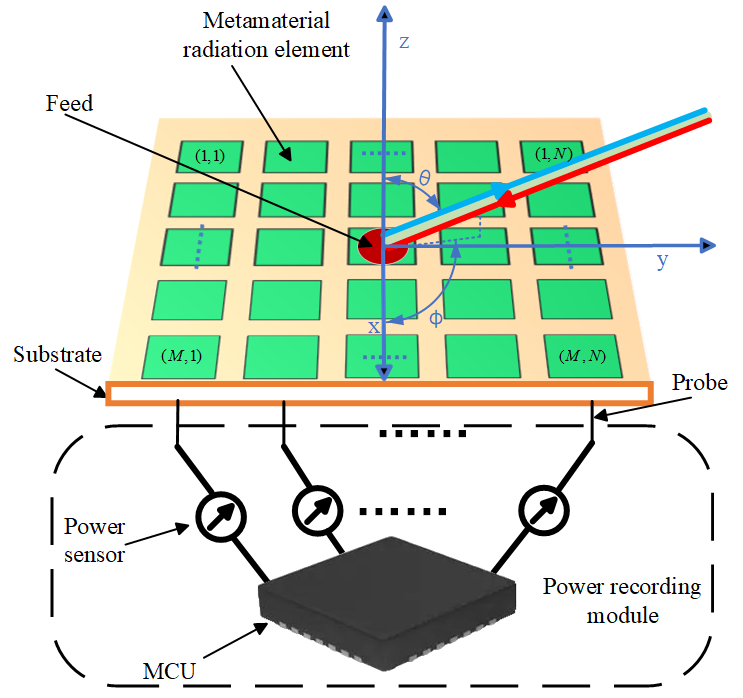}
       \caption{\rev{ Illustration of the structure of RRM.} }
       \label{fig2}
\end{figure}

\subsection{Recording}

As shown in Fig. 3(a), a reference wave is fed to the surface, and it interferes with the received waveform from the user, forming a hologram in Fig. 3(b). The intensity of the interference is measured by the power sensors and recorded. Next, we elaborate the recording process.



    

\begin{figure}[!ht]
    \centering
    \begin{minipage}{0.48\linewidth}
        \centering
        \includegraphics[width=\linewidth]{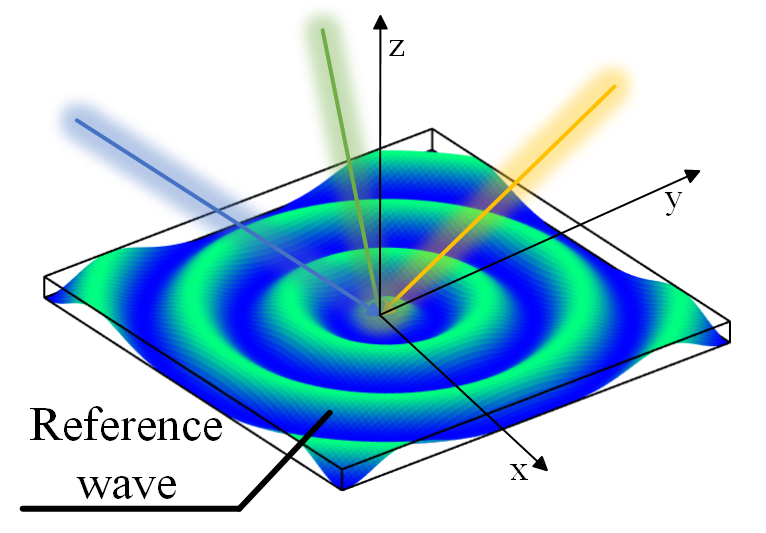}
        \scriptsize{(a)}
        \label{fig3a}
    \end{minipage}
    \hspace{1mm}
    \begin{minipage}{0.48\linewidth}
        \centering
        \includegraphics[width=\linewidth]{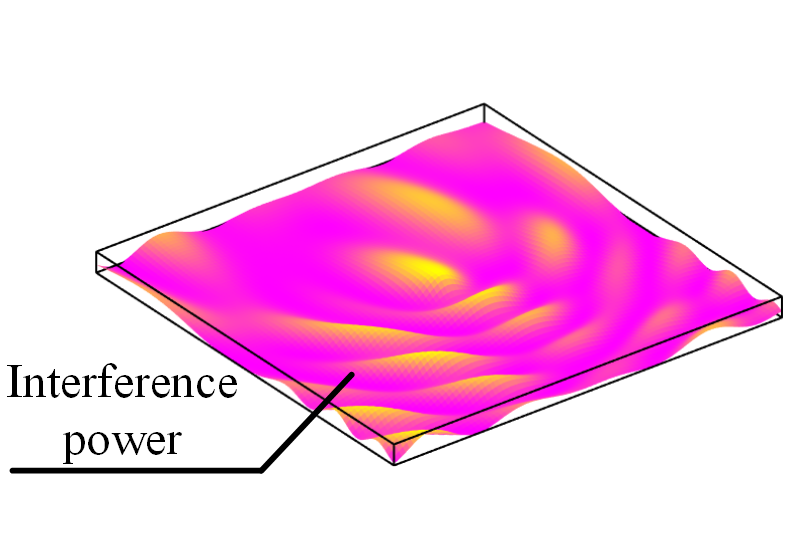}
        \scriptsize{(b)}
        \label{fig3b}
    \end{minipage}
    \caption{\rev{Illustration of the interference in EM holography: (a) the reference wave interferes with the object waves; (b) Interference power on RRM.}}
    \label{fig:main}
\end{figure}



We assume that the RRM consists of $M\times N$ elements and the feed is located at the geometric center. At the $(m, n)$-th element of the surface, the reference wave can be represented as
\begin{equation}
\label{eq5}
{E}_{{r}}(m, n)=A_{r} e^{\mathrm{j} k_{r} d_{r}(m,n) },   
\end{equation}
where $A_{r}$ and $k_{r}$ respectively represent the amplitude and wave number of the reference wave when it propagates through the substrate, $d_{r}(m,n) = \sqrt{\left[d_{x}(m-(M+1)/2)\right]^{2}+\left[d_{y}(n-(N+1)/2)\right]^{2}}$ denotes the distance between the feed and the $(m, n)$-th radiation element, and $d_{x}$ and $d_{y}$ represent the element spacing in the $x$ and $y$ axes on the RRMs, respectively. At the $(m, n)$-th element, the superimposed field composed of $N_{p}$ object waves incident on the RRMs from directions $\{(\theta_{p},\phi_{p})\}$ can be expressed as 
\begin{equation}
\label{eq6}
 \begin{aligned}
E_{o}(m, n) &= \sum_{p=1}^{N_{p}} E_{o}^{p}(m, n) \\
&= \sum_{p=1}^{N_{p}} A_{p} e^{\mathrm{j}( k_{f} d_{o}^{p}(m, n)+\varphi_{o}^{p})},
\end{aligned}   
\end{equation}
where {$A_{p}$ and $\varphi_{o}^{p}$ represent the amplitude and initial phase of the $p$-th object wave incident from direction ($\theta_{p},\phi_{p}$), and $d_{o}^{p}(m,n) =d_{x}(m-1) \sin \theta_{p} \cos \phi_{p}+d_{y}(m-1) \sin \theta_{p} \sin \phi_{p} $ denotes the distance between the geometric center and the $(m,n)$-th antenna unit on the RRMs.}  The interference between the reference wave and object waves can be expressed as 
\begin{equation}
    {E}_{{c}}(m, n)={E}_{{o}}(m, n)+{E}_{{r}}(m, n).
\end{equation}
It is noted that the {interference power at the metasurface contains the information required for our subsequent holographic beamforming, 
which is measured by a power sensor integrated within each surface element. The power distribution of the interference across the surface, resulting in a hologram shown in Fig. 3(b).} The \({W}(m,n) \) in the interference power matrix \(  \boldsymbol{W} \) is given by
\begin{equation}
  \begin{aligned}
{W}(m, n) =&{E}_{{c}}(m, n) {E}_{c}^{*}(m, n) \\
=&\left|{E}_{{o}}(m, n)\right|^{2}+\left|{E}_{{r}}(m, n)\right|^{2}\\
&+{E}_{{o}}(m, n) {E}_{{r}}^{*}(m, n)\\
&+{E}_{{o}}^{*}(m, n) {E}_{{r}}(m, n).
\end{aligned}  
\end{equation}


\subsection{Holographic Beamforming (Reconstruction)}
Similar to RHSs, instead of using phase shifting, RRMs adopt an amplitude controlling method to construct the holographic pattern. The power of the leaky-wave antennas on the RRM is determined by the hologram, i.e., the leaky-wave antennas at different locations have different output powers when excited by the reference wave, achieving the effect of beamforming. {The question we need to answer is how to obtain the holographic weights to achieve transmit beamforming in the opposite directions of the object waves.} 

\rev{As shown in Fig. 2}, we adopt a Cartesian coordinate system, where the $x$-$y$ plane coincides with the RRM, the $z$-axis is perpendicular to the RRM, and the origin is at the geometric center of the RRM. In this context, the elements of the interference power matrix \( \boldsymbol{W} \) correspond one-to-one with the antenna elements. It can be observed that the index center of the matrix \(((M+1)/2, (N+1)/2)\), coincides with the origin of the coordinate system. Under the plane wave assumption, the electromagnetic waves received and transmitted in opposite directions are represented by the same azimuth and elevation angles. 

We now investigate the relationship between EM waves with opposite directions. In free space, the complex amplitude of an electric field plane wave can be expressed as
\begin{equation}
{E}(\boldsymbol{r}) = A_{0} e^{\mathrm{j} \boldsymbol{k} \cdot \boldsymbol{r}},
\end{equation}
where \( A \) is the amplitude of the electric field, \( \boldsymbol{k} \) is the wave vector in free space, representing the propagation direction of the electric field, and \( \boldsymbol{r} \) is the spatial position vector. Therefore, when the propagation direction of the electromagnetic wave is reversed, the direction of the wave vector also reverses. That is, the reverse electric field propagates along \(-\boldsymbol{k}\). As a result, there exists a conjugate relationship between the reverse-propagating electric field and the original electric field, which can be denoted by \( {E}^*(\boldsymbol{r}) =  A_{0} e^{- \mathrm{j}\boldsymbol{k} \cdot \boldsymbol{r}} \). From the above, we need the presence of the conjugate term \( E_o^*(m,n)\) in the reconstructed object wave.

    
The following definition provides a method for matrix reindexing, which offers a convenient way to obtain the holographic weights.

\newtheorem{definition}{Definition}
\begin{definition}
For arbitrary matrix \( \boldsymbol{W} \in \mathbb{C}^{M \times N} \), the reindexed matrix \( \boldsymbol{W}^{\prime} \) is obtained by the following element-wise mapping: $W^{\prime}(m, n) = W(M - m + 1, N - n + 1)$, for \( m = 1, 2, \dots, M \) and \( n = 1, 2, \dots, N \). 
\end{definition}


With the help of \rev{Fig. 2}, it can be observed that the matrix reindexing method is to perform a 180-degree rotation of the matrix elements around the z-axis. Then, this method can be applied to reindex the interference power matrix $\boldsymbol{W}$, thereby obtaining the holographic weight matrix $\boldsymbol{W}^{\prime}$ that generates the desired beam direction.

\newtheorem{proposition}{Proposition}
\begin{proposition}
Assume that the antenna elements of the RRM are arranged in a centrosymmetric configuration, and the feed source is located at the geometric center of the RRM. The elements of the holographic weight matrix $W^{\prime}(m, n) = W(M - m + 1, N - n + 1)$ are obtained by reindexing the interference power matrix \( \boldsymbol{W} \) during the recording process.
Through the holographic weight matrix \( \boldsymbol{W}^{\prime} \), the RRM can generate a reconstructed object wave as
\begin{subequations}
\begin{equation}
\begin{aligned}
{E}_{{h}}(m, n)= & {E}_{{r}}(m, n) {W}^{\prime}(m, n) \\
=& A_c {E}_{{r}}(m, n)+\left|{E}_{{r}}(m, n)\right|^{2} {E}_{{o}}^{*}(m, n)\\    
&+\sum_{i=1}^{N_{p}} \sum_{j=1}^{N_{p}} {E}_{{o}}^{{i}}(m, n)\left[{E}_{{o}}^{{j}}(m, n)\right]^{*}{E}_{{r}}(m, n)\\
&+{E}_{{o}}(m, n) \left [ {E}_{{r}}(m, n)\right]^{2} ,i \neq j, 
\end{aligned}
\end{equation}
\begin{equation}
\begin{aligned}
    A_c &= \left|{E}_{{r}}(m, n)\right|^{2}+\sum_{p=1}^{N_{p}}\left|{E}_{{o}}^{p}(m, n)\right|^{2}\\
    &= A_{r}^{2}+\sum_{p=1}^{N_p} A_{p}^{2}, 
\end{aligned}    
\end{equation}
\end{subequations}
where the term $\left|{E}_{{r}}(m, n)\right|^{2} {E}_{{o}}^{*}(m, n)$ represents the component propagating in the opposite directions to the incident object wave.
\end{proposition}


\textit{Proof:} See Appendix A.




{We note that the direct use of the holographic weights $\boldsymbol{W}^{\prime}$ for holographic beamforming is not efficient. This can be seen from (10a) that only the term $\left|{E}_{{r}}(m, n)\right|^{2} {E}^{*}_{o}(m, n)$ is useful for beamforming in desired directions, and all other terms lead to sidelobes, causing interference to signals in the desired directions.}
{It can be observed that the intensity variations caused by different terms due to the reference wave amplitude are not identical. Except the first term, the second and fourth terms are proportional to the square of the reference wave amplitude. Therefore, by appropriately choosing a larger reference wave amplitude, the proportion of the third term in the overall interference power can be reduced. However, a relatively large reference wave amplitude will significantly increase the proportion of the first term.}

{To improve the beamforming efficiency, we can also deal with the constant term by subtracting an appropriate constant $b$ from the holographic weight matrix \( \boldsymbol{W}^{\prime} \), so that the impact of the constant term $A_c$ is best mitigated while keeping the weights are positive numbers. Since the energy of the incoming wave is unknown, the value of $b$ can be set to the square of the reference wave amplitude. Thus, we have}
\begin{equation}
{W}_{h}(m,n)= \rho\left({W}^{\prime}(m,n)-b\right).   
\end{equation}
For example, the value of $b$ can take the average of the elements of \( \boldsymbol{W}^{\prime} \). Noting that the subtraction may lead to negative values, we can simply force the negative values to be zero. 
The normalization factor $\rho$ restricts the overall values of the matrix elements to the range $[0,1]$.

\subsection{Comparison with Conventional Beamforming Approaches }

\begin{figure*}[t]
  \centering
  \includegraphics[width=0.85\textwidth]{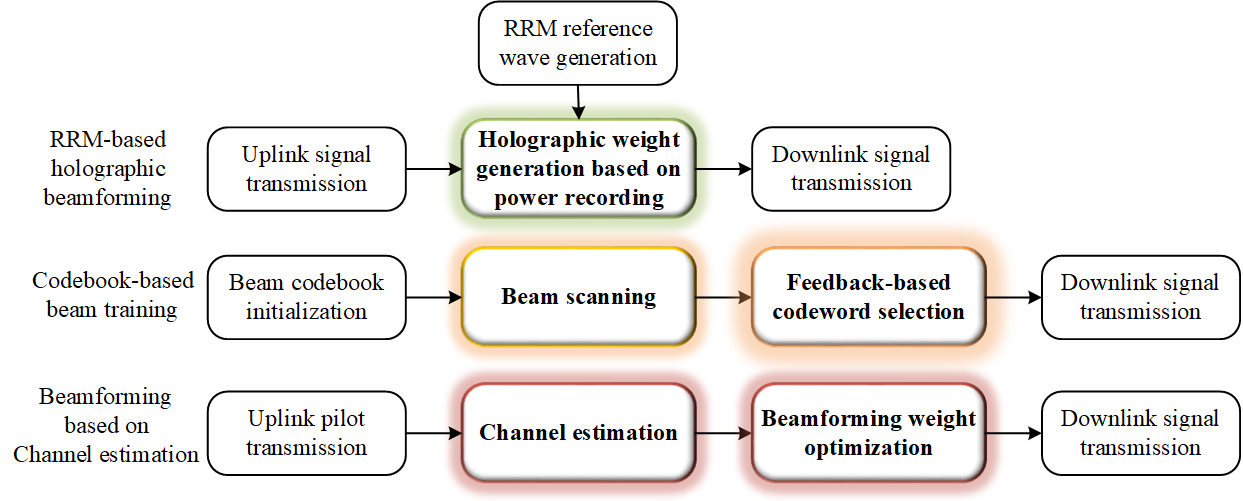}
  \caption{\rev{RRM-based Holographic Beamforming Scheme vs. Conventional Beamforming Scheme}}
  \label{fig:beamforming-compare}
\end{figure*}

\rev{Following the introduction of the recording and reconstruction in RRM-based holographic beamforming, we present Fig. \ref{fig:beamforming-compare} to visually highlight the differences between our proposed scheme and conventional beamforming approaches. For comparison, we have selected two classic beamforming schemes: beamforming based on channel estimation\cite{ref33,ref34} and beam training based on codebook\cite{ref35,ref36} .}

\rev{In Fig. \ref{fig:beamforming-compare}, we use different colors to emphasize the core steps of each scheme. Our proposed RRM-based holographic beamforming method sidesteps the need for beam scanning and CSI acquisition by directly recording the spatial interference power and then computing holographic weights that steer energy toward the desired direction. The interference power measurement yields an exceptionally lightweight implementation in both computation and overhead, rendering it highly suitable for large-scale antenna arrays. In contrast, the codebook-based beam training scheme relies on user feedback and high-overhead beam scanning, resulting in significant consumption of time-frequency resources. The scheme based on channel estimation uses uplink pilot symbols to estimate the CSI before deriving beamforming weights. While this method can achieve near-optimal performance, the high computational complexity affects its application in large-scale arrays.}

\rev{In comparison, our proposed solution only needs to record interference power within a short period and involves extremely low-complexity computations to perform beamforming. The recording process essentially stores spatial phase information through power measurements, giving it distinct advantages in various communication scenarios that satisfy channel reciprocity.}

\subsection{Relation to RHS-Based Holographic Beamforming }\label{3B}


{In the RHS-based scheme, there is no recording process, so no physical hologram is formed. It assumes that the CSI and beamforming directions (object wave) are available at the transmitter side, so that the hologram term useful for beamforming is directly computed. This can avoid the non-useful terms in the hologram that contribute to sidelobes, thereby achieving better performance. To enable this, the knowledge of object wave at the surface is needed. Subsequently, we will demonstrate the working principle of the RHS by using the example of a single desired beamforming direction.}

{The reference wave on the RHS can also be represented by (\ref{eq5}). 
{At the $(m,n)$-th radiation element, the wave to achieve the desired propagation direction $(\theta_{0},\phi_{0})$ is given by}}
\begin{equation}
E_{d}(m, n) = A_{d} e^{-\mathrm{j} k_{f} d_{o}^{0}(m, n)}, 
\end{equation}
{Hence the useful hologram term for beamforming is}
\begin{equation}
{W}_{\text{int}}(m, n) = {E}_{{d}}(m, n){E}_{{r}}^{*}(m, n).
\end{equation}
The holographic pattern excited by the reference wave can be expressed as
\begin{equation}
{E}_{t}(m, n) = {E}_{{d}}(m, n)\left|{E}_{{r}}(m, n)\right|^{2}.
\end{equation}
Thus, the object wave radiating in the desired direction $(\theta_{0},\phi_{0})$ can be generated. 
{It is noted that the beamforming weights described above is  complex-valued.} {The radiation elements that emit waves in phase with the desired directional beam are adjusted to maximize their radiation intensity, whereas the elements that are out of phase are detuned to minimize their radiation output.} However, RHS controls the radiation intensity of the reference wave at radiating elements to perform beamforming. Given the above reasons, it is necessary to adjust the {weights}. Since the real part of the weight represents the cosine value of the phase difference between the object wave and the reference wave, it can be used for amplitude control. 
Even though the RHS-based scheme directly computes the useful components affecting the phase in interference, it still needs to satisfy the constraint that the holographic matrix elements are positive and real-value constraint. Therefore, the holographic weights will also contain interference terms.
{Assume that $\operatorname{Re}\left[\boldsymbol{W}_{i n t}\right]$ is normalized to $[0,1]$, then the positive valued holographic weight is given as \cite{ref18}}
\begin{equation}
\begin{aligned}
{M}\left(m,n\right)=&\frac{\operatorname{Re}\left[W_{\text{int}}\left(m,n\right)\right]  +1}{2}\\ 
=&\frac{1}{2}+\frac{1}{4}{E}_{r}(m, n){E}_{d}^{*}(m, n)\\
&+\frac{1}{4}{E}_{r}^{*}(m, n){E}_{d}(m, n). 
\end{aligned}
\end{equation}


\begin{table*}[t]
\begin{center}
\caption{Comparison of RRMs and RHS.}
\label{tab1}
\begin{tabular}{| c | c | c |}
\hline
 & RHS & RRMs\\
\hline
Hardware Components& Feed, Substrate, Antenna elements & Feed, Substrate, Antenna elements, {Power sensor}\\
\hline
& Using conventional methods to obtain CSI,  &
 Obtain interference power through the {recording process}, \\ 
Working Principle&  compute holographic weights based on & and compute holographic weights with \\ 
& the desired direction.& low complexity operations.\\
\hline
\rev{Computational Overhead} & \rev{High} & \rev{Low}\\
\hline
\end{tabular}
\end{center}
\end{table*}

\section{Holographic Communications Using RRMs}
\label{sec:4}

{In this section, with our RRM-based holographic beamforming, we derive the communicatin system model and analyze its mutual information. 
We consider a TDMA communication system with multipath propagation 
As shown in Fig.~1(c) and (d), the user is equipped with a single antenna and the RRM has size $M\times N$. \rev{Since the RRM-based holographic scheme uses information recorded during the uplink to synthesize downlink beamforming, we assume that the propagation channel is slowly varying and satisfies reciprocity. As channel reciprocity is a result of the Rayleigh–Carson reciprocity theorem, it is further assumed that the properties of the transmission medium and scatterers are symmetric for both the uplink and downlink \cite{ref27}. Empirical validations of reciprocity at millimeter-wave frequencies on practical MIMO and phased-array platforms have also been reported \cite{ref37,ref38}.}


\subsection {Recording and Holographic Weights Determination}
{As shown in Fig. 1 (c), the recording takes place during the uplink transmission. } 
{The passband signal} with amplitude $A_{u}$ and angular frequency $w_{c}$ transmitted by the user can be represented by
\begin{equation}
s(t)=\sqrt{2} \mathrm{Re}\left(A_{u} e^{\mathrm{j} \omega_{c} t}\right).
\end{equation}
{It is noted that an information-carrying phase-modulated signal can be used for the recording process. As the extension is straightforward, the discussion is omitted here.}
The channel impulse response from the user to the ($m$,$n$)-th {element of the surface can be expressed as}
\begin{equation}
{h}_{m,n}(t)=\sum_{i=1}^{L} \alpha^{i} {a}_{m,n}\left(\theta^{i}, \phi^{i}\right) \delta\left(t-\tau_{i}\right), 
\end{equation}
where $\alpha_{i}$ represents the gain coefficient of the $i$-th path, $\tau_{i}$ denotes the time delay of the $i$-th path, $\delta\left( \cdot \right)$ is the Dirac delta function and {${a}_{m,n}\left(\theta^{i}, \phi^{i}\right)$ is the  the $mn$-th element of the steering vector $a(\theta^{i},\phi^{i})$} along the direction specified by ($\theta^{i},\phi^{i}$) for the $i$-th path incident on the RRMs. 
{The geometric center of the RRMs is chosen as the reference
point of the steering vector. Then the element ${a}_{m,n}^{i}\left(\theta^{i}, \phi^{i}\right)$} is given as
\begin{subequations}
\begin{equation}
{a}_{m,n}^{i}\left(\theta^{i}, \phi^{i}\right)= \mathrm{e}^{-\mathrm{j}k_{f} d_{m,n}\left(\theta^{i}, \phi^{i}\right)},
\end{equation}
\begin{equation}
\begin{aligned}
 d_{m,n}\left(\theta^{i}, \phi^{i}\right)= &d_{x}\left(m-\frac{M+1}{2}\right) \sin \theta_{i} \cos \phi_{i}\\
 &+d_{y}\left(n-\frac{N+1}{2}\right) \sin \theta_{i} \sin \phi_{i}.   
\end{aligned} 
\end{equation}
\end{subequations}
The signal received by the $(m, n)$-th antenna of the RRM can be expressed as
\begin{equation}
    y_{m, n}^{u}(t)=\sqrt{2} \mathrm{Re}\left[h_{m, n}(t) * A_{u} e^{\mathrm{j} \omega_{c} t}+z(t)\right],
\end{equation}
where $z(t) \sim \mathcal{CN}(0, \sigma^2)$ is the additive white Gaussian noise (AWGN). Assuming {that there exits} a phase difference $\varphi$ between the {recording} reference signal and the reference signal in the subsequent reconstruction phase mentioned below, the generation of a local reference signal with amplitude $A_{r}$ and angular frequency $\omega_{r}$ is given by
\begin{subequations}
 \begin{equation}
   y_{m, n}^{r}(t)=\sqrt{2} \mathrm{Re}\left[A_{r} \mathrm{e}^{\mathrm{j}\left(w_{r} t+\phi\right)} \beta_{m, n}\right], 
\end{equation}
\begin{equation}
\beta_{m,n}= e^{-\mathrm{j} k_{r} \sqrt{\left[d_{x}(m-(M+1) / 2)\right]^{2}+\left[d_{y}(n-(N+1) / 2)\right]^{2}}},
\end{equation}   
\end{subequations}
{where ${\beta_{m.n}}$ represents the relative phase shift of the reference wave at the $(m, n)$-th antenna element} when the feed is located at the geometric center of RRMs. It is required that 
$\omega_{r}=\omega_{c}$ for interference.
Thus, {the interference signal at RRMs can be represented as} 
\begin{equation}
y_{m, n}^{c}(t)=y_{m, n}^{r}(t)+y_{m, n}^{u}(t). 
\end{equation}
\begin{subequations}
\begin{figure*}[ht] 
\centering 
\begin{equation}
\label{formula: bounds on mu}
\begin{aligned}
W_{m, n}(t)= & \left|\sum_{i=1}^{L} A_{u} e^{\mathrm{j} \omega_{r}\left(t-\tau_{i}\right)} \alpha^{i} a_{m, n}^{i}\left(\theta^{i}, \phi^{i}\right) +A_{r}\mathrm{e}^{\mathrm{j}\left(w_{r} t+\varphi\right)} \beta_{m, n}+z(t)\right|^{2}\\
= & A+A_{r} A_{u} \mathrm{e}^{\mathrm{j} \varphi} \beta_{m,n} \sum_{i=1}^{L}\left[\alpha^{i} a_{m, n}^{i}\left(\theta^{i}, \phi^{i}\right) e^{-\mathrm{j} \omega_{r} \tau_{i}}\right]^{*}+A_{r} A_{u} \mathrm{e}^{-\mathrm{j} \varphi} \beta_{m, n}^{*} \sum_{i=1}^{L}\left[\alpha^{i} a_{m_{z}, n}^{i}\left(\theta^{i}, \phi^{i}\right) e^{-\mathrm{j} \omega_{r} \tau_{i}}\right]\\ 
& +\sum_{i=1}^{L} \sum_{j=i+1}^{L} 2 A_{u}^{2} \mathrm{Re}\left\{\left[\alpha^{i} a_{m, n}^{i}\left(\theta^{i}, \phi^{i}\right) e^{-\mathrm{j} \omega_{r} \tau_{i}}\right]\left[\alpha^{j} a_{m, n}^{j}\left(\theta^{j}, \phi^{j}\right) e^{-\mathrm{j} \omega_{r} \tau_{j}}\right]^{*}\right\}+|z(t)|^{2}+n(t) ,
\end{aligned}
\end{equation}
\end{figure*}
Then we obtain the interference power of the local reference signal and the user signal at the $(m, n)$-th element by (\ref{formula: bounds on mu}), where

\begin{equation}
A=A_{r}{ }^{2}+\sum_{i=1}^{L}\left|\alpha^{i}\right|^{2} A_{u}{ }^{2},    
\end{equation}
\begin{small}
 \begin{equation}
n(t) = 2 \mathrm{Re}\left\{ z(t) \left[
\begin{aligned}
& \sum_{i=1}^{L} A_{u} \alpha^{i} a_{m, n}\left(\theta^{i}, \phi^{i}\right) e^{-\mathrm{j} \omega_{r} \tau_{i}} \\
& + A_{r} e^{\mathrm{j}\left(\omega_{r} t + \varphi\right)} \beta_{m, n}
\end{aligned}
\right]^{*} \right\}.
\end{equation}     
\end{small}
\end{subequations}

\begin{figure*}[ht] 
\centering 
\begin{equation}
\label{con:inventoryflow}
\begin{aligned}
{W}_{m, n}^{\prime}(t)= & A+A_{r} A_{u} \mathrm{e}^{-\mathrm{j} \varphi} \beta_{m, n} \sum_{i=1}^{L} a_{m, n}^{*}\left(\theta^{i}, \phi^{i}\right)\left[\alpha^{i} e^{-\mathrm{j} \omega_{r} \tau_{i}}\right]+A_{r} A_{u} \mathrm{e}^{\mathrm{j} \varphi} \beta_{m, n} \sum_{i=1}^{L} a_{m, n}\left(\theta^{i}, \phi^{i}\right)\left[\alpha^{i} e^{-\mathrm{j} \omega_{r} \tau_{i}}\right]^{*} \\
& +\sum_{i=1}^{L} \sum_{j=i+1}^{L} 2 A_{u}^{2} \mathrm{Re}\left\{a_{m, n}^{*}\left(\theta^{i}, \phi^{i}\right)\left[\alpha^{i} e^{-\mathrm{j} \omega_{r} \tau_{i}}\right] a_{m, n}\left(\theta^{j}, \phi^{j}\right)\left[\alpha^{j} e^{-\mathrm{j} \omega_{r} \tau_{j}}\right]^{*}\right\}+|z(t)|^{2}+n(t).
\end{aligned}
\end{equation}
\hrulefill 
\vspace*{4pt} 
\end{figure*}
{Then according to Proposition 1, after reindexing the obtained interference power matrix $\boldsymbol{W}$, we obtain the holographic weights in (\ref{con:inventoryflow}). After that, we can subtract a constant term from \( W_{m,n}^{\prime} \) as discussed in Section \ref{sec:3} to suppress the sidelobes, resulting in the holographic weight \( W_{m,n}^{h} \).} 

\subsection{Signal Model for Downlink}
{With the holographic weights, holographic beamforming can be performed and downlink transmission can be implemented through feeding a communication signal to the surface. It is noted that the reference signal is a communication signal with phase and amplitude modulation (which has the same frequency as the reference signal during the recording process), so that information can be transmitted.} 

The information sequence to be transmitted with a length of $k$ is denoted as {$u[1], ...u[k]$}. Then the baseband complex signal $x_{b}(t)$ is given by
\begin{equation}
x_{b}(t)=\sum_{n=0}^{k} u[n] \delta\left(t-n T_{s}\right),
\end{equation}
where $T_{s}$ is the symbol duration. After pulse shaping with a filter $q(t)$, we obtain the baseband signal
\begin{equation}
\tilde{x}_{b}(t) =\sum_{n=0}^{k} u[n] p(t-n T_{s}).
\end{equation}
{Thus, we have the modulated signal that is fed to the surface
\begin{equation}
y_{m, n}^{R}(t)=Re [A_{r} \tilde{x}_{b}(t) \mathrm{e}^{\mathrm{j} w_{r} t}]. 
\end{equation}
With the holographic weight matrix $\boldsymbol{W}^{h}$, 
the radiated signal corresponding to the $(m,n)$-th element can be expressed as
\begin{equation}
\tilde{y}_{m, n}(t)=\sqrt{2} W_{m, n}^{h} y_{m, n}^{R}(t).
\end{equation}}

Assume that the channel is quasi-static across the recording phase and downlink transmission phase using. After passing through a reciprocal channel, the signal received by the user 
is given as
\begin{small}
\begin{equation}
\mathrm{Re}\left[\tilde{\boldsymbol{y}}_{b}(t) \mathrm{e}^{\mathrm{j} w_{r} t}\right]=\mathrm{Re}\left[\sum_{m, n} h_{m, n}(t) *\left[W_{m, n}^{h} \tilde{x}_{b}(t) y_{m, n}^{R}(t)\right]+\boldsymbol{z}(t)\right].
\end{equation}
\end{small}

Thus, the baseband received signal $\tilde{\boldsymbol{y}}_{b}(t)$ obtained after down-conversion can be expressed as
\begin{subequations}
\begin{equation}
\tilde{y}_{b}(t) = \rho \left\{
\begin{aligned}
 M& N A_{r}^{2} A_{u} \mathrm{e}^{\mathrm{j} \varphi} \sum_{i=1}^{L} (\alpha^{i})^{2} \mathrm{e}^{-\mathrm{j} \omega_{r} 2 \tau_{i}} \tilde{x}_{b}(t-\tau_{i}) \\
 +& \sum_{m, n} \sum_{i=1}^{L} p_{m, n}^{i}(A_{r} W_{m, n}^{e} \beta_{m, n} \\
& + A_{r}^{2} A_{u} \mathrm{e}^{\mathrm{j} \varphi} \sum_{j=1, j \neq i}^{L} p_{m, n}^{j}) \tilde{x}_{b}(t-\tau_{i})
\end{aligned}
\right\} + z(t)
\end{equation}
with
\begin{equation}
\begin{aligned}
W_{m, n}^{e} =  A& + A_{r} A_{u} \mathrm{e}^{-\mathrm{j} \varphi} \beta_{m, n} \sum_{i=1}^{L} a_{m, n}\left(\theta^{i}, \phi^{i}\right)\left[\alpha^{i} e^{-\mathrm{j} \omega_{r} \tau_{i}}\right]^{*} \\
 +& \sum_{i=1}^{L} \sum_{j=i+1}^{L} 2 \operatorname{Re}\left\{a_{m, n}\left(\theta^{i}, \phi^{i}\right)^{*}\left[\alpha^{i} e^{-\mathrm{j} \omega_{r} \tau_{i}}\right] \right.\\
& \left.\times a_{m, n}\left(\theta^{j}, \phi^{j}\right)\left[\alpha^{j} e^{-\mathrm{j} \omega_{r} \tau_{j}}\right]^{*}\right\} - b ,
\end{aligned}
\end{equation}
\begin{equation}
p_{m, n}^{i}=\alpha^{i} a_{m, n}^{i}\left(\theta^{i}, \phi^{i}\right) \mathrm{e}^{-\mathrm{j} \omega_{r} \tau_{i}},
\end{equation}
\end{subequations}
where $\rho$ is the normalization factor, $W_{m, n}^{e}$ represents the residual terms obtained after removing the major contribution term from the holographic weights. The baseband signal after matched filtering with filter $q(t)$ is given by
\begin{subequations}
 \begin{equation}
\begin{aligned}
y_{b}(t) & =q(t) * \tilde{y}_{b}(t) \\
& =h_{b}(t) * w(t) * x_{b}(t)+q(t) * z(t),
\end{aligned}    
\end{equation}
where
\begin{equation}
h_{b}(t)=\sum_{i=1}^{L} \alpha_{h}^{i} \delta\left(t-\tau_{i}\right),
\end{equation}
\begin{equation}
\alpha_{h}^{i}=\alpha_{h 1}^{i}+\alpha_{h 2}^{i} \ ,
\end{equation}
\begin{equation}
  \alpha_{h 1}^{i}=\rho M N A_{r}{ }^{2} A_{u} \mathrm{e}^{\mathrm{j} \varphi}\left(\alpha^{i}\right)^{2} \mathrm{e}^{\mathrm{j} \omega_{r} 2 \tau_{i}}, 
\end{equation}
\begin{equation}
    \small{\alpha_{h 2}^{i}=\rho \sum_{m, n} \sum_{j=1, j \neq i}^{L} p_{m, n}^{i}\left(A_{r}{ }^{2} A_{u} \mathrm{e}^{\mathrm{j} \varphi} p_{m, n}^{j}+A_{r} I_{m, n}^{e} \beta_{m, n}\right)}.  
\end{equation}   
\end{subequations}
In particular, the equivalent channel coefficient $\alpha_{h 1}^{i}$ of the $i$-th path can be split into $\alpha_{h 1}^{i}$ and $\alpha_{h 2}^{i} $, where $\alpha_{h 1}^{i}$ is the main contribution term, which includes the response of main beaming, and $\alpha_{h 2}^{i} $ represents other terms. Finally, the continuous-time baseband signal is sampled at a rate of $1/T_{s}$, resulting in a discrete-time signal $y_{b}[m]$ as
\begin{subequations}
 \begin{equation}
 \label{y=hx}
 \begin{aligned}
y_{b}[m]=&h_{b}\left(m T_{s}\right) * w\left(m T_{s}\right) * x_{b}\left(m T_{s}\right)\\
 &+q\left(m T_{s}\right) * z\left(m T_{s}\right) \\
 =&\sum_{n=0}^{k-1} u[n] \sum_{i=1}^{L} \alpha_{h}^{i} w\left(m T_{s}-n T_{s}-\tau_{i}\right)\\
&+\sum_{j=0}^{k-1} q\left(j T_{s}\right) z\left(m T_{s}-j T_{s}\right) \\
 =&\sum_{n=0}^{k-1} u[n] \sum_{i=1}^{L} \alpha_{h}^{i} w\left[m-n-\frac{\tau_{i}}{T_{s}}\right]\\
 &+\sum_{j=0}^{k-1} q[j] z[m-j] \\
 =&\sum_{l=m}^{m-k+1} h[l] u[m-l]+\sum_{j=0}^{k-1} q[j] z[m-j],
\end{aligned} 
\end{equation}

\begin{equation}
   h[l]=\sum_{i=1}^{L} \alpha_{h}^{i} w\left[l-\frac{\tau_{i}}{T_{s}}\right], l=m-n.
\end{equation}       
\end{subequations}
We can rewrite (\ref{y=hx}) in a vector form as  
\begin{subequations}
 \begin{equation}
\boldsymbol{y}=\boldsymbol{H}\boldsymbol{u}+\boldsymbol{F}\boldsymbol{z},
\end{equation}
\begin{equation}
\boldsymbol{y}=\left[\begin{array}{llll}
y_{b}[0] & y_{b}[1] & \cdots & y_{b}[k-1]
\end{array}\right]^{\mathrm{T}},
\end{equation}   
\end{subequations}
where $\boldsymbol{u}$ and $\boldsymbol{z}$ represent the transmitted signal vector and noise vector, respectively. 
The channel matrix $\boldsymbol{H}\in \mathbb{C}^{K \times K}$ is a Toeplitz matrix formed by the channel coefficients $h[l]$ at different time instants, {as shown in (\ref{H})}.
Matrix $\boldsymbol{F}\in \mathbb{C}^{K \times K}$ is a Toeplitz matrix, constructed using the impulse response $q(t)$ of the RRC filter. 
 When root raised cosine pulse shaping is used with symbol sampling period, the noise is white, i.e., 
$\boldsymbol{F}$ is the identity matrix.

\begin{figure*}[ht] 
\centering 
\begin{equation}
\label{H}
\boldsymbol{H} = \setlength{\arraycolsep}{2pt} 
\begin{bmatrix}
    h[0] & h[-1] & \ldots & h[-L+1] & 0 & 0 & \ldots & 0 & 0 \\
    h[-1] & h[0] & \ldots & h[-L] & h[-L+1] & 0 & \ldots & 0 & 0 \\
    \vdots & \vdots & \ddots & \vdots & \vdots & \vdots & \ddots & \vdots & \vdots \\
    0 & 0 & \ldots & 0 & 0 & h[L-1] & \ldots & h[1] & h[0]
\end{bmatrix}.
\end{equation}
\vspace*{4pt} 
\hrulefill 
\end{figure*}

{Based on the established model, the input-output mutual information $\mathcal{I}(\boldsymbol{u};\boldsymbol{y})$ can be expressed as}
\begin{equation}
 \begin{aligned}
\mathcal{I}(\boldsymbol{u};\boldsymbol{y})&=\frac{1}{K}\log _{2}\left[\operatorname{det}\left(\boldsymbol{I}+\frac{\boldsymbol{H} \boldsymbol{Q} \boldsymbol{H}^{\mathrm{H}}} {N_{0}} \right)\right] \\
&=\frac{1}{K}\log _{2}\left[\operatorname{det}\left(\boldsymbol{I}+\gamma\boldsymbol{H} \boldsymbol{H}^{\mathrm{H}}\right)\right],
\end{aligned}   
\end{equation}
{where $Q$ is the covariance matrix of the transmitted signal. With the i.i.d. Gaussian assumption,  
$Q=\gamma{N_{0}}\boldsymbol{I}$,} where \(\gamma\) represents the average signal-to-noise ratio (SNR) at the receiver, and \(\boldsymbol{I}\) denotes a \(K \times K\) identity matrix. Furthermore, we can derive the outage probability $P_{out}$ of the system at the threshold rate $R_{th}$ as follows
\begin{equation}
P_{\mathrm{out}}(R_{th}) = \Pr \left\{ \mathcal{I}(\boldsymbol{u};\boldsymbol{y}) < R_{th} \right\}
\end{equation}

\rev{In the following simulation experiments, we will use mutual information and outage probability as performance metrics to fully evaluate the performance of the RRM-based holographic communication system.}

\section{Simulation Results}
\label{sec:5}
In this section, we evaluate the performance of the proposed RRM-based holographic communication scheme. We assume one line-of-sight (LOS) path and four nonline-of-sight (NLOS) paths between the user and RRM. 
\begin{figure}[H]
    \centering
    \subfloat[{Beam pattern of the RRM without dealing with the constant term.}]{ 
        \includegraphics[width=0.38\textwidth]{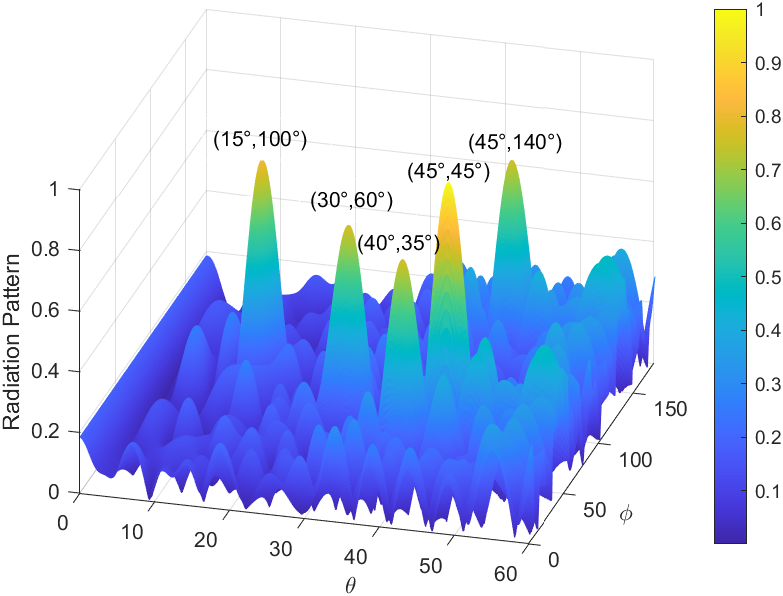}
        \label{BEAMa}
    }\\
    \subfloat[{Beam pattern of the RRM with the constant term considered.}]{ 
        \includegraphics[width=0.38\textwidth]{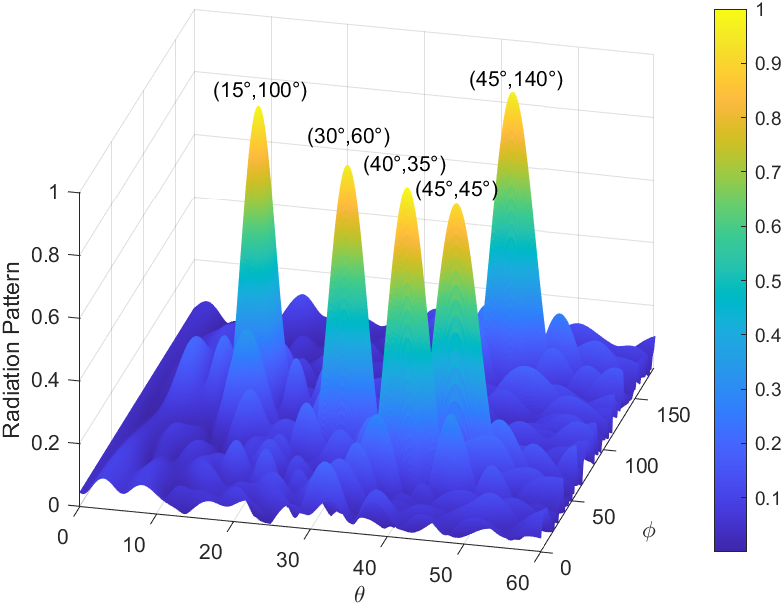}
        \label{BEAMb}
    }\\
    \caption{{Beam pattern of the RRM with the holographic weight obtained from recording.}}
    \label{fig:beamPattern}
\end{figure}
Simulation parameters are set according to the 3GPP specifications \cite{ref25}.
The carrier frequency $f_c$ is 30 GHz. $k_{f} = 2\pi f_c/c$ is the wavenumber in the free space, and $k_{s}=\sqrt{3} k_{f}$ is the wavenumber in the RRM, where $c$ is the speed of light. The element spacing of the RRM and RHS (for comparison) in both directions ($d_{x}$ and $d_{y}$) is half-wavelength. The transmit power of the RRM and RHS $P_{T}$ is 1 W. As mentioned early, the RHS-based communication scheme with the knowledge of perfect CSI in \ref{3B} is compared with the proposed RRM-based scheme. 

\begin{figure}[H]
\centering
\subfloat[RRM size is $8 \times 8$.]{
		\includegraphics[width=0.45\textwidth]{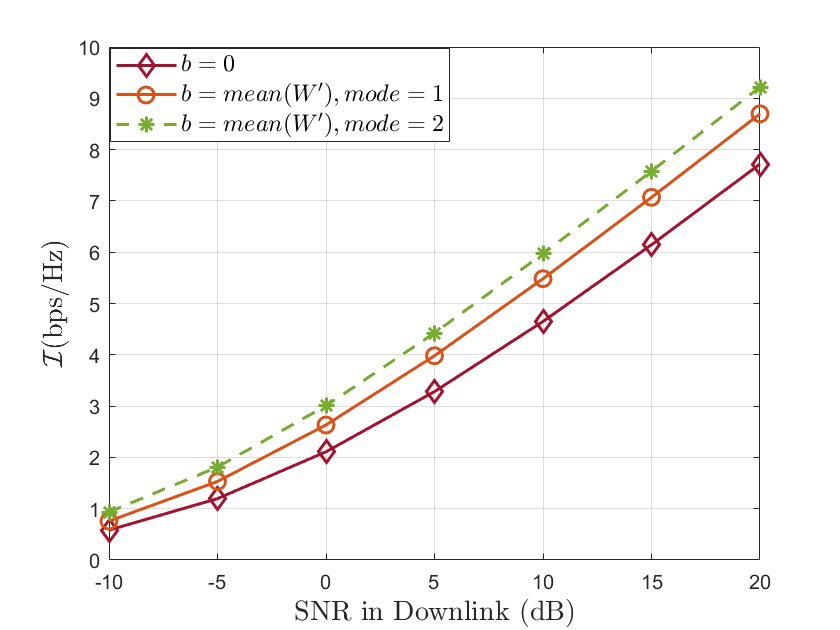}
        \label{b-8}
}\\
\subfloat[RRM size is $64\times 64$.]{
		\includegraphics[width=0.45\textwidth]{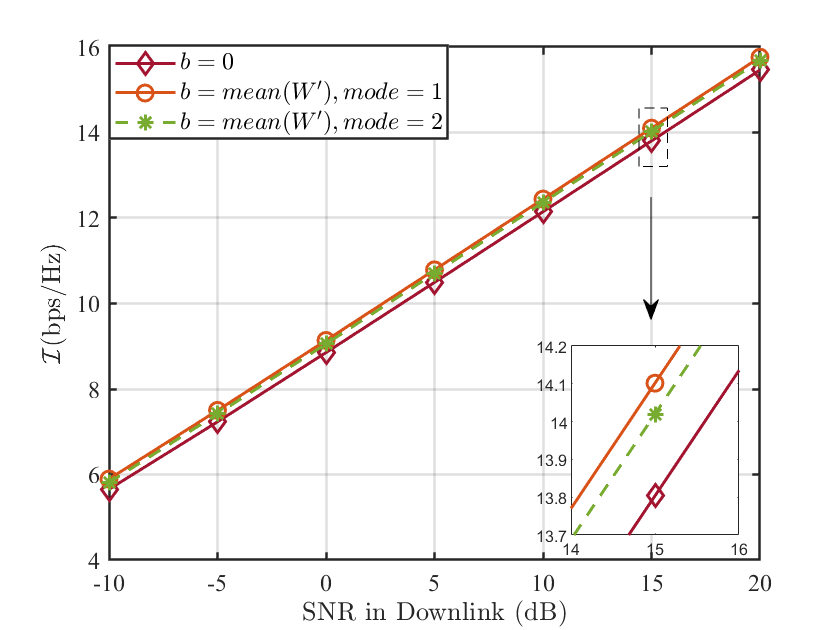}
        \label{b-64}
}\\
\caption{Mutual information versus SNR with different values of $b$.}
\label{figb}
\end{figure}

{We first evaluate the beamforming performance, where the holographic weights are obtained through the recording process.} {The RRM size is set to 32×32.} 
{We assume that the user's multipath signals incident on the RRM with angles 
$(15^\circ ,100^\circ), (30^\circ ,60^\circ), (40^\circ,35^\circ), (45^\circ,45^\circ)$, and $(45^\circ,140^\circ)$. Fig. \ref{fig:beamPattern} illustrates the normalized beam pattern of RRM using the holographic weights.} Specifically, Fig. \ref{fig:beamPattern}\subref{BEAMa} presents the results without dealing with the influence of the constant term, i.e., $b=0$ {in (11)}, while in Fig. \ref{fig:beamPattern}\subref{BEAMb} a constant term $b$, which is the average of the elements of $\boldsymbol{W}^{\prime}$ i.e., \( b=\text{mean}(\boldsymbol{W}^{\prime}) \), is subtracted from the original weights. This may lead to negative values, which are then set to 0. 
We see that the directions of the beams generated by the RRMs match well with the directions of the paths. 
{This demonstrates that the RRM can perform effective beamforming without CSI using the proposed scheme. Moreover, comparing Fig. \ref{fig:beamPattern}\subref{BEAMb} with Fig. \ref{fig:beamPattern}\subref{BEAMa}, it is evident that dealing with the constant term can effectively suppress the influence of sidelobes, thereby enhancing the beamforming performance.}


To quantify the gain introduced by the constant $b$, Fig. \ref{figb} illustrate the impact of different values of $b$ on the mutual information with a varying size of RRM. To address the potential issue of negative values that may arise after subtracting \( b \) from the elements of the holographic weights, we propose two approaches to selecting \( b \). In Approach 1, consistent with the Beam Pattern simulation, \( b \) is set as \( b=\text{mean}(\boldsymbol{W}^{\prime}) \), and any resulting negative values are set to zero. In Approach 2, \( b \) is set to the minimum value of the elements in \( \boldsymbol{W}^{\prime} \), 
thereby avoiding the occurrence of negative values.

By comparing Fig. \ref{figb}\subref{b-8} and Fig. \ref{figb}\subref{b-64}, it can be observed that the gain from dealing with the constant term is more pronounced when the size of the RRM is smaller. This is because, at smaller sizes, the beams are relatively wide, the sidelobes generated by the constant term generates more interference to the main beams. Therefore, dealing with the constant term leads to more significant gains. 
 {Furthermore, when the size is larger, the performance gain from both methods decreases due to the stronger beam directivity, resulting in a convergence of performance characteristics between the two methods.} For smaller sizes, dealing with the constant term with \( b = \text{mean}(\boldsymbol{W}^{\prime}) \) is closer to the theoretical true value of the constant term and yields better results. {As the size increases, the interference from the constant term has a reduced impact on beamforming}, and the holographic weight matrix constructed by this approach contains more zero values, 
which negatively impacts the beamforming performance.

   \begin{figure}[H]
        \centering
        \includegraphics[width=0.45\textwidth]{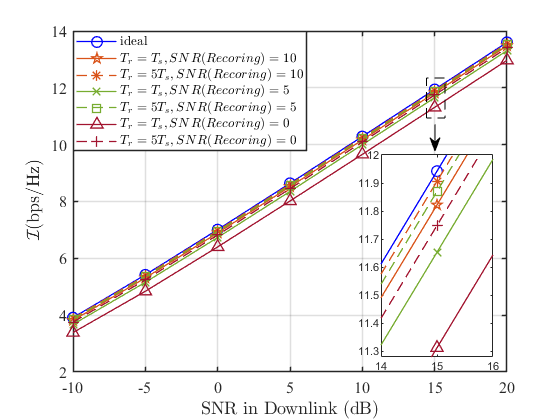}
		\caption{Mutual information versus SNR with various recording durations.}
        \label{rs}
    \end{figure}

Fig. \ref{rs} illustrates the mutual information versus SNR and the recording time, where the RRM has  a dimension of $32 \times 32$. 
It can be seen from Fig. \ref{rs} that, {the SNR during recording has a relatively small effect on the quality of holographic downlink communication. This is because, during the recording process, the thermal noise at the antenna elements has the same power. 
Therefore, the impact of noise attributes to a constant term in the interference power, having a small effect on the quality of holographic beamforming. Furthermore, we also examine the impact of recording duration. 
It is shown that increasing the recording duration can further improve the mutual information. This is because, with a longer recording duration, the noise power measured by the power meters at all antenna elements is closer to the true noise power. 
This reduces the impact of randomness in noise power estimation and makes the noise power at different antenna elements be a same constant.}

{Fig. \ref{RR}  shows the mutual information versus SNR, where the size of the RRM is $8\times8$, $16\times16$, $32\times32$, and $64\times64$.} 
{We assume that the RHS-based scheme has perfect CSI, while the RRM completes the recording process under an SNR of 10 dB with \( T = 5T_s \).} It can be seen that, the mutual information increases with the SNR and the size of the RRM.
This is because, as the size of the antenna array increases, the holographic beamformer achieves more focused beam directivity and higher power utilization efficiency, leading to an increase in mutual information. In addition, at smaller sizes, the gain from dealing with the constant term is more pronounced, leading to a performance advantage for the RRM compared to the RHS. At larger sizes, the performance of the RRM remains very close to that of the RHS. The simulation results also demonstrate that, even under noisy recording conditions, the RRM still exhibits performance comparable to that of the RHS with perfect CSI.
   \begin{figure}[H]
        \centering
        \includegraphics[width=0.45\textwidth]{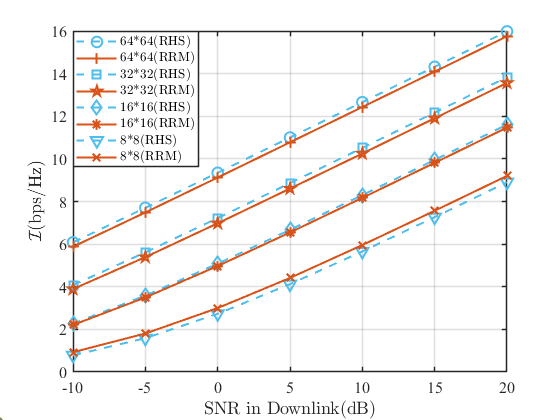}
		\caption{ Mutual information versus SNR with a varying size of RRM.}
        \label{RR}
    \end{figure}

\rev{Fig. \ref{CDL} evaluates the proposed RRM scheme under the more realistic CDL-D channel model, showing the mutual information versus SNR for various sizes. While the mutual information increases with both SNR and array size as expected, the performance of RRM relative to RHS in Fig. \ref{CDL} is slightly lower than that in Fig. \ref{RR}. This mild performance difference is primarily due to the CDL-D channel’s richer delay spread and angular dispersion which increase the proportion of non-useful interference terms in the recorded hologram and thus yield larger residual sidelobes after reconstruction. Nevertheless, the RRM scheme remains close to the RHS baseline with perfect CSI. }

   \begin{figure}[H]
        \centering
        \includegraphics[width=0.45\textwidth]{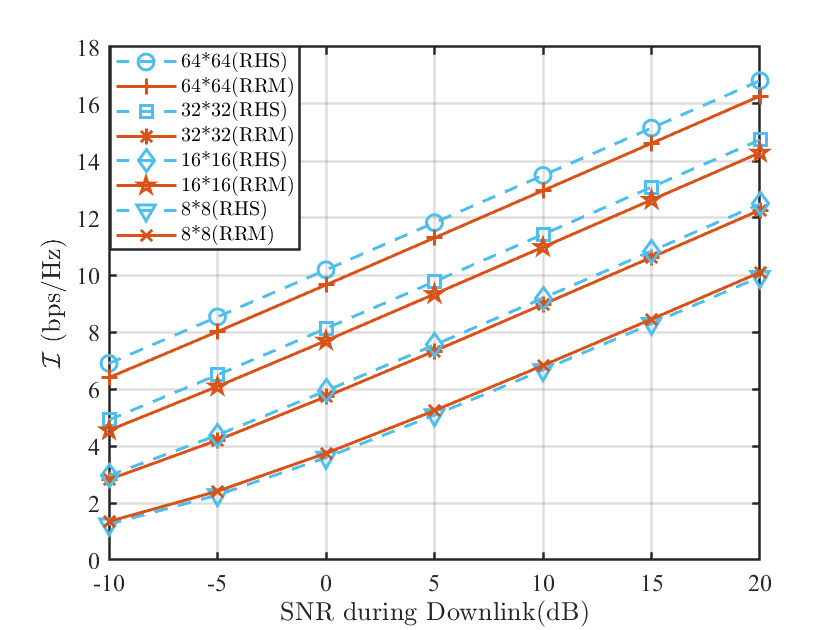}
		\caption{ \rev{Mutual information versus SNR with varying RRM size under the CDL-D channel.}}
        \label{CDL}
    \end{figure}

\rev{Fig. \ref{OP}  illustrates the outage probability versus SNR, where the size of the RRM is $8\times8$ and $16\times16$. It can be observed that when the threshold rate $R_{th}=2$, the system outage probability of RRM is lower than that of RHS under the $8\times8$ antenna configuration, demonstrating a certain performance advantage. However, under the $16\times16$ antenna configuration, the outage probability of RRM becomes marginally higher than that of RHS. This indicates that the trend of performance variation aligns with that shown in Fig. \ref{RR}, and provides a more pronounced and comprehensive reflection of the performance gap between RRM and RHS.}  

   \begin{figure}[H]
        \centering
        \includegraphics[width=0.45\textwidth]{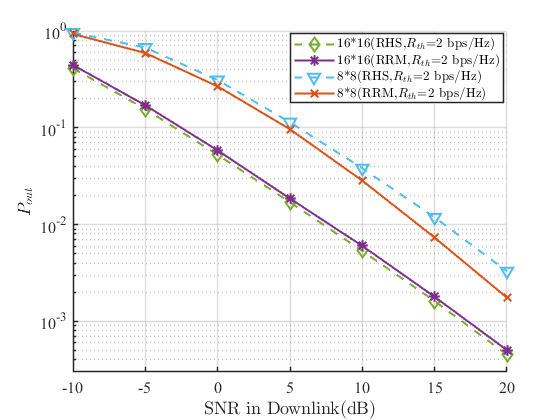}
		\caption{ \rev{Outage Probability versus SNR with a varying size of RRM.}}
        \label{OP}
    \end{figure}

\section{Conclusions}
\label{sec:6}
{In this paper, we proposed a new EM holographic communication scheme using RRMs, which involves a recording process and a holographic beamforming process. With the recording process enabled by RRMs, the power of the interference between the user waveform and reference waveform is obtained, based on which the holographic weights for beamforming can be determined. Thus, the challenging channel estimation is bypassed. We also derived the signal model for the RRM-based communication system and analyze its mutual information. We showed that the proposed RRM-based system performs similarly to the RHS-based system with perfect CSI, providing a promising alternative for metasurface based communications. In this work, we focus on a TDMA system. 
Our future work includes the extension of this work to the case of multiple user concurrent communications.}        


{\appendix[Proof of Proposition 1]
The interference power $W(m,n)$ described in (8) can be rewritten as
\begin{equation}
  \begin{aligned}
{W}(m, n) =&\sum_{p=1}^{N_{p}}\left|{E}_{{o}}^{p}(m, n)\right|^{2}+\left|{E}_{{r}}(m, n)\right|^{2}\\
&+{E}_{{o}}(m, n) {E}_{{r}}^{*}(m, n)\\
&+{E}_{{o}}^{*}(m, n) {E}_{{r}}(m, n)\\
&+\sum_{i=1}^{N_{p}} \sum_{j=1}^{N_{p}} {E}_{{o}}^{{i}}(m, n)\left[{E}_{{o}}^{{j}}(m, n)\right]^{*} ,i \neq j. 
\end{aligned}  
\end{equation}

Subsequently, we reindex matrix \( \boldsymbol{W} \) to obtain \( \boldsymbol{W}^{\prime} \). It is noted that, as the first and second terms in the equation are constants, they will keep unchanged after the reindexing operation. Then, after the reindexing, we have

\begin{equation}
  \begin{aligned}
{W}^{\prime}(m, n) =&\sum_{p=1}^{N_{p}}\left|{E}_{{o}}^{p}(m, n)\right|^{2}+\left|{E}_{{r}}(m, n)\right|^{2}\\
&+{E}_{{o}}^{\prime}(m, n) {E}_{{r}}^{{\prime}^{*}}(m, n)\\
&+{E}_{{o}}^{{\prime}^{*}}(m, n) {E}_{{r}}^{\prime}(m, n)\\
&+\sum_{i=1}^{N_{p}} \sum_{j=1}^{N_{p}} {{E}_{{o}}^{i}}^{\prime}(m, n)\left[{{E}_{{o}}^{j}}^{\prime}(m, n)\right]^{*} ,i \neq j
\end{aligned}  
\end{equation}
where \( {E}_o^{\prime}(m, n) \) and \( {E}_r^{\prime}(m, n) \) are the elements of the matrices obtained by reindexing \( \boldsymbol{E}_o \) and \( \boldsymbol{E}_r \), respectively. 
By the expression of the object wave in (6), it can be seen that
\begin{equation}
\label{eq36}
 \begin{aligned}
E_{o}^{\prime}(m, n)&= E_{o}(M+1-m, N+1-n)\\
&= \sum_{p=1}^{N_{p}} A_{p} e^{\mathrm{j} k_{f} d_{o}^{p}(M+1-m, N+1-n)} ,
\end{aligned}   
\end{equation}
where
\begin{equation}
\begin{aligned}
d_{o}^{p}(&M+1-m, N+1-n)\\
= &\quad d_{x}(M+1-m-(M+1) / 2)\sin \theta_{p} \cos \phi_{p}\\
&+d_{y}(N+1-n-(N+1) / 2) \sin \theta_{p} \sin \phi_{p}\\ 
=&-d_{x}(m-(M+1) / 2)\sin \theta_{p} \cos \phi_{p}\\
&-d_{y}(n-(N+1) / 2) \sin \theta_{p} \sin \phi_{p}\\ 
=&-d_{o}^{p}(m, n).
\end{aligned}
\end{equation}
Therefore, \({E}_{{o}}^{\prime}(m, n)\) can be further expressed as
\begin{equation}
\begin{aligned}
E_{o}^{\prime}(m, n)
&= \sum_{p=1}^{N_{p}} A_{p} e^{-\mathrm{j} k_{f} d_{o}^{p}(M+1-m, N+1-n)} \\
&=E_{o}^{*}(m, n),
\end{aligned}   
\end{equation}
It can be observed that the matrix reindexing method proposed allows the conjugate of the complex amplitude of the object wave at the RRM antenna position to be obtained. Regarding the reference wave, 
${E}_{{r}}^{\prime}(m, n)$ can be expressed as
\begin{equation}
\begin{aligned}
{E}_{{r}}^{\prime}(m, n)=&{E}_{{r}}(M+1-m, N+1-n)\\
=&A_{r} e^{\mathrm{j} k_{r} d_{r}(M+1-m, N+1-n) }\\
=&A_{r} e^{\mathrm{j} k_{r} d_{r}(m, n) }\\
=&{E}_{{r}}(m, n).
\end{aligned}   
\end{equation}
Since the reference wave is excited by the feed, and the feed is located at the geometric center of the RRM, the complex amplitude of the reference wave at the RRM antenna position is unaffected by the matrix reindexing. Substituting the results obtained from (39) and (40) into (36), we have
\begin{equation}
  \begin{aligned}
{W}^{\prime}(m, n) =&\sum_{p=1}^{N_{p}}\left|{E}_{{o}}^{p}(m, n)\right|^{2}+\left|{E}_{{r}}(m, n)\right|^{2}\\
&+{E}_{{o}}^{*}(m, n) {E}_{{r}}^{*}(m, n)\\
&+{E}_{{o}}(m, n) {E}_{{r}}(m, n)\\
&+\sum_{i=1}^{N_{p}} \sum_{j=1}^{N_{p}} {{E}_{{o}}^{i}}(m, n)\left[{{E}_{{o}}^{j}}(m, n)\right]^{*} ,i \neq j. 
\end{aligned}  
\end{equation}

We see that that matrix \( \boldsymbol{W}^{\prime} \) contains the desired term \( {E}_{{o}}^{*}(m, n) {E}_{{r}}^{*}(m, n) \), which can thus serve as the holographic weights for beamforming in the desired directions. By loading the holographic weights $\boldsymbol{W}^{\prime}$ obtained from the aforementioned process onto the antenna elements and exciting them using the reference wave, we obtain the reconstructed object wave as in (10).

 }


\newpage


\end{document}